\documentclass[12pt]{article}

\usepackage[numbers,sort&compress]{natbib}
\usepackage{graphicx}
\usepackage{epsfig}
\usepackage{amsmath}
\usepackage{amssymb}
\usepackage{hyperref}

\usepackage{multimedia}
\usepackage{pgf}

\usepackage{tikz}
\usetikzlibrary{arrows}
\usetikzlibrary{trees,snakes}
\tikzstyle{block}=[draw opacity=0.7,line width=1.4cm]

\newcommand{\beq}{\begin{equation}}
\newcommand{\eeq}{\end{equation}}
\newcommand{\beqs}{\begin{eqnarray}}
\newcommand{\eeqs}{\end{eqnarray}}
\newcommand{\Tr}{\ensuremath{\mathop{\mathrm{Tr}}}}

\def\ni{\noindent}
\def\be{\begin{equation}}
\def\ee{\end{equation}}
\def\bea{\begin{eqnarray}}
\def\eea{\end{eqnarray}}
\def\bsp{\be\begin{split}}
\def\la{\langle}
\def\ra{\rangle}

\def\wt{\widetilde}

\def\G{\Gamma}

\def\S{\Sigma}
\def\a{\alpha}

\def\g{\gamma}

\def\d{\delta}
\def\e{\epsilon}
\def\m{\mu}

\def\n{\nu}
\def\s{\sigma}

\def\l{\lambda}
\def\t{\tau}
\def\o{\omega}
\def\O{\Omega}
\def\T{\theta}

\def\p{\partial}

\def\bR {\mathbb{R}}

\def\bX {\mathbb{X}}
\def\w{\wedge}

\makeatletter

\newcommand{\Rmnum}[1]{\expandafter\@slowromancap\romannumeral #1@}
\makeatother

\renewcommand{\title}[1]{\vbox{\center\LARGE{#1}}\vspace{5mm}}
\renewcommand{\author}[1]{\vbox{\center\large{#1}}\vspace{5mm}}
\newcommand{\address}[1]{\vbox{\center\em#1}}
\newcommand{\email}[1]{\vbox{\center\tt#1}\vspace{5mm}}

\begin{document}
\bibliographystyle{utphys}

\begin{titlepage}
\begin{flushright}
{\tt AEI-2009-036}\\
{\tt HU-EP-09/15}
\end{flushright}
\title{\vspace{1.0in} {\bf Supersymmetric Wilson Loops in Diverse Dimensions}}

\author{Abhishek Agarwal$^1$ and Donovan Young$^2$}

\address{$^1$Max-Planck-Institut f\"ur Gravitationsphysik\\
  Albert-Einstein-Institut\\
  Am M\"uhlenberg 1, D-14476 Potsdam, Germany\\
  $^2$Humboldt-Universit\"at zu Berlin, Institut f\"ur Physik,\\
  Newtonstra\ss e 15, D-12489 Berlin, Germany }

\email{$^1$abhishek@aei.mpg.de, $^2$dyoung@physik.hu-berlin.de}

\abstract{\ni We consider supersymmetric Wilson loops \`a la Zarembo
  in planar supersymmetric Yang-Mills theories in diverse
  dimensions. Using perturbation theory we show that these loops have
  trivial vacuum expectation values to second order in the 't Hooft
  coupling. We review the known superspace results which, for specific
  dimensions, extend this triviality to all orders in the 't Hooft
  coupling. Using the gauge/gravity correspondence, we construct the
  explicit dual fundamental string solutions corresponding to these
  Wilson loops for the case of circular geometry. We find that the
  regularized action of these string solutions vanishes. We also
  generalize the framework of calibrated surfaces to prove the
  vanishing of the regularized action for loops of general
  geometry. We propose a possible string-side manifestation of the
  gauge theory generalized Konishi anomaly in seven dimensions.}

\end{titlepage}


\section{Introduction}

By now there is mounting evidence in favor of both the usefulness as
well as the validity of the gauge/gravity duality between $\mathcal{N}
=4 $ SYM in four dimensions and string theory on $AdS_5\times S^5$. It
is thus natural to ask if this duality can be tested and utilized in
the cases of gauge theories in dimensions other than four. Such
investigations are naturally motivated by the need to understand how
the gauge/gravity duality may be realized in non-conformal
supersymmetric Yang-Mills theories. For the special cases of sixteen
supercharge SYM theories in diverse dimensions, the gravity duals were
proposed some time ago in \cite{Itzhaki:1998dd}. While the feasibility
of generic tests of gauge/gravity duality is not very clear for
sixteen supercharge SYM theories in dimensions greater than four: the
SYM theories are not renormalizable and the dual D$p$-brane geometries
suffer from the non-decoupling of the alpha-prime corrections, it is
worthwhile to exploit the duality between SYM theories in $p+1$
dimensions and D$p$-branes and test it in the case of protected
operators whose vacuum expectation values are independent of the
coupling $g^2$.  A special class of Wilson loops, first proposed by
Zarembo in the case of $\mathcal{N}=4$ SYM \cite{Zarembo:2002an} are
particularly well suited to this purpose. In this paper we generalize
Zarembo's construction to perform a non-trivial test of the duality
between D$p$-brane theories and SYM in $p+1$ dimensions.

The Maldacena-Wilson loop \cite{Maldacena:1998im,Rey:1998ik} has proven to be a
very powerful probe of the AdS/CFT correspondence. In four dimensional
${\cal N}=4$ supersymmetric Yang-Mills theory it is given by
\be\label{wl}
W = \frac{1}{N} \Tr P \exp \oint d\t\,\Bigl( i\dot x^\m(\t) A_\m + |\dot x(\t)|
\Theta^I(\t) \Phi^I \Bigr),
\ee
where $N$ is the rank of the gauge group $SU(N)$, $\Phi^I$ are the six
scalar fields of the theory, and $\Theta^I\Theta^I=1$. The path of the
Wilson loop is defined by $x^\m(\t)$, but there is also the freedom to
define a path on $S^5$ parametrized by $\Theta^I(\t)$. The specific
coupling to the scalar fields in (\ref{wl}) is chosen to ensure local
supersymmetry; the amount of global supersymmetry respected by $W$ is
intimately connected with the correlation of the paths $x^\m(\t)$ and
$\Theta^I(\t)$. There is a ``perfectly'' correlated choice, found by
Zarembo \cite{Zarembo:2002an}\footnote{These Wilson loops are closely
  related to a class constructed later in
  \cite{Drukker:2007yx,Drukker:2007dw,Drukker:2007qr} whose contours
  lie on a three-sphere.}
\be\label{pc}
\Theta^I (\t) =  \frac{\dot x^\m}{|\dot x|}\,M^I_\m, \quad M^I_\m M^I_\n
= \d_{\m\n},
\ee
where $M^I_\m$ is a constant matrix, which assures that the vacuum
expectation value of the Wilson loop is trivial
\be\label{weq1}
\la W \ra_{\text{Zarembo}} = 1.
\ee
The amount of supersymmetry respected by the loop is found by requiring
\be\label{gtss}
\d_\e W \sim \dot x^\m \Bigl( i\g_\m + M^I_\m \G_I \Bigr) \e = 0.
\ee
This gives one halving of the supersymmetry\footnote{The Poincar\'e
  and superconformal supersymmetries are halved independently of one
  another.} for each non-zero component of $\dot x^\m$, so that, for
example, a planar loop is $1/4$ BPS. One can appreciate the result (\ref{weq1})
from a few different perspectives. The first is that (\ref{pc})
ensures that the combined gauge and scalar field Feynman gauge
propagator joining two points on the loop is zero
\bsp
&\left\la \Bigl( i\dot x^\m(\t) A_\m + |\dot x(\t)|
\Theta^I_x(\t) \Phi^I \Bigr) \Bigl( i\dot y^\n(\s) A_\n + |\dot y(\s)|
\Theta^J_y(\s) \Phi^J \Bigr)
\right\ra\\
&\qquad\qquad\qquad\qquad\qquad\qquad\qquad\qquad\qquad
=\frac{g^2}{4\pi^2}\, \frac{-\dot x \cdot \dot y + \Theta_x \cdot \Theta_y |\dot
  x||\dot y|}{(x-y)^2} = 0,
\end{split}
\ee
which immediately precludes the contribution of ladder/rainbow
diagrams.  As shown by Zarembo \cite{Zarembo:2002an}, all interacting
diagrams up to two loops can also, without an inordinate effort, be
shown to vanish.  A much stronger statement was made in
\cite{Guralnik:2003di}, where superspace techniques were exploited to
prove (\ref{weq1}) for Wilson loops whose contours are contained in
$\bR^3$. The loops of Zarembo are also naturally described in terms of
the twisting of ${\cal N}=4$ SYM to produce a topological theory; in
this context the triviality of the vacuum expectation value for loops
in the full $\bR^4$ was proven in \cite{Kapustin:2006pk}.

At strong coupling the vacuum expectation value of (\ref{wl}) is
accessible via the dual string theory. It is given by the partition
function of a fundamental string, the saddle points of which are
minimal area embeddings in $AdS_5 \times S^5$
\cite{Maldacena:1998im,Drukker:1999zq}. In the following coordinates
for $AdS_5 \times S^5$
\be
ds^2 = U^2 dX^\m dX^\m + \frac{1}{U^2} dU^I dU^I,
\ee
one requires the following boundary conditions for the string
embedding $\S$ at the boundary $U= \infty$
\be\label{gbc}
X^\m |_{\p \S} = x^\m,\qquad \left.
\frac{U^I}{|U|} \right|_{\p\S} = \Theta^I.
\ee
The action of the string is then found to contain a generic divergence
owing to the diverging area element of Anti-de Sitter space as the
boundary is approached. This divergence is proportional to the
circumference of the loop
\be
S = \frac{\sqrt{\l}}{2\pi}\int d^2\s \sqrt{\det \p_a \bX^M \p_b \bX^N G_{MN}} =
\frac{\sqrt{\l}}{2\pi}
\left( U_{\text{max.}} \oint d\t |\dot x(\t)| + A_{\text{reg.}}\right),
\ee
where $\bX^M=(X^\m,U^I)$, and may be removed via a Legendre
transformation \cite{Drukker:1999zq}, leaving the regularized action
$S_{\text{reg.}}=\sqrt{\l}A_{\text{reg.}}/(2\pi)$. The result for the
vacuum expectation value of (\ref{wl}) is then
\be
\la W \ra_{\l \to \infty} = {\cal V}\,\exp\left(-S_{\text{reg.}}\right),
\ee
where ${\cal V}$ is a prefactor stemming from integration over zero
modes in the partition function. The disc partition function naturally
involves three zero modes. If there is no extra parametric freedom in
embedding the string, then ${\cal V} \sim \lambda^{-3/4}$, i.e. one
factor of $\lambda^{-1/4}$ for each zero mode. This is the case for
the standard $1/2$ BPS circle which sits at a point on $S^5$
\cite{Drukker:2000rr}.  The expectation therefore, for the string dual
of the Zarembo loops, is that ${\cal V}=1$, and
$S_{\text{reg.}}=0$. The first of these conditions has not been shown
explicitly, and for other than planar loops remains a mystery. For the
case of planar loops, it was argued in \cite{Zarembo:2002an} that
there are 3 compensating zero modes stemming from parametric freedom
in embedding the string in an $S^2 \subset S^5$. For loops other than
planar, it remains unclear how the contribution of the three basic
zero modes is cancelled \cite{Dymarsky:2006ve}. We discuss this issue
further in section \ref{sec:zm}. The second condition,
$S_{\text{reg.}}=0$, was shown explicitly by Zarembo in
\cite{Zarembo:2002an} for the circular supersymmetric Wilson loop in
$AdS_5\times S^5$. There the string solution was found and the
regularized action calculated. The analogous string-side embodiment of
the results of \cite{Guralnik:2003di} were realized\footnote{The issue
  of the zero mode prefactor ${\cal V}$ is still outstanding.} in
\cite{Dymarsky:2006ve}, where it was proven that $S_{\text{reg.}}=0$
for the string dual of a generic ${\cal N}=4$ SYM Zarembo Wilson
loop. This used the method of calibrated surfaces which we will review
in section \ref{sec:cal}.

In the present work we will extend these results, to the degree it is
possible, to maximally supersymmetric Yang-Mills theories in general
spacetime dimensions. Indeed we may view (\ref{wl}) as arising from a
toroidal compactification of the standard Wilson loop in ${\cal N}=1$,
$d=10$ SYM, and in this sense we are free to compactify more or less
directions than 6, namely $9-p$ where $p$ ranges from 0 to 9,
\bsp
\frac{1}{N} \Tr P \exp &\oint d\t\, i\dot x^M(\t) A_M \\
&\to \frac{1}{N} \Tr P \exp \oint d\t\,\Bigl( i\dot x^\m(\t) A_\m +
|\dot x(\t)|
\Theta^I(\t) \Phi^I \Bigr) ,
\end{split}
\ee
where $M=1,\ldots,10$, $\m = 1,\ldots,p+1$, $I=1,\ldots,9-p$. We then
require the same relations to hold relating the paths $x^\m$ and
$\Theta^I$, i.e. (\ref{pc}). The supersymmetry relation (\ref{gtss})
also continues to hold after this dimensional reduction. For the
various spacetime dimensions $d$, we are restricted by (\ref{pc}) to
curves $x^\m(\t)$ in various subspaces of $\bR^d$, these are
summarized in the table below.
\begin{center}
\begin{tabular}{c|ccccccccc}
  \hline
  $d$ & 1 &  2 & 3 & 4 & 5 & 6 & 7 & 8 & 9\\
  Curves in & $\bR^1$ & $\bR^2$ & $\bR^3$ & $\bR^4$ & $\bR^5$ &
  $\bR^4$ & $\bR^3$ & $\bR^2$ & $\bR^1$\\
  \hline
\end{tabular}
\end{center}
We will concentrate on the dimensions $2 \leq d \leq 8$, since
the curves in $\bR^1$ are the trivial 1/2 BPS straight lines.

On the gauge theory side, we perform our analyses using both
perturbative and (non-perturbative) superspace techniques. From the
perturbative point of view, we study the relevant gauge theories in a
unified way, up to the next to leading order (NLO), or two loop
approximation. This analysis allows us to perform a straightforward
extension of the results presented in \cite{Zarembo:2002an}. At this
order in perturbation theory we find that the vacuum expectation value
for the Zarembo loops in {\it all} dimensional reductions of the
$d=10$, $\mathcal{N}=1$ SYM theories, down to $d=1$, is identically `1'.
Clearly, the NLO results beg the question if some or all of the gauge
theories preserve the triviality of the Zarembo loops to higher or
even all orders in perturbation theory.

On a related note, one may also worry about the reasonability of
perturbative methods in non-renormalizable gauge theories, which SYM
in $d\geq 5$ are expected to be. Though we do not expect the
perturbative results for generic gauge theory observables in these
theories to be meaningful, we can use perturbation theory to gauge the
validity of results believed to be protected by non-renormalization
theorems. The non-renormalization theorems for the sixteen supercharge
theories in question were derived in \cite{Guralnik:2003di}. In that
paper, the $d$ dimensional SYM theories were reformulated in a $d-3$
superspace language. This reformulation, which is briefly reviewed in
the next section, allows one to view the Wilson loops in question as
elements of a chiral ring. Furthermore, the (superspace) equations
of motion were shown to imply shape invariance of the loops embeddable
in $\bR^3$.  These two results were used to formally establish the
triviality of these Zarembo loops for all sixteen supercharge gauge
theories in $7> d \geq 3$. The appearance of a generalized Konishi anomaly in
$d=7$ \cite{Guralnik:2003di} puts an upper bound (in terms of
dimensions) on the gauge theories for which the perturbative results
may be expected to hold to all loop orders. However, for gauge
theories in $d <3$, the superspace methods are simply limited by the
construction/requirement of a $d-3$ dimensional superspace, with at
least one dynamical supercoordinate. We can thus regard the
perturbative results as a non-trivial verification of the predictions
of \cite{Guralnik:2003di} at the NLO, and a hint toward the potential
for generalization of the triviality of the Zarembo loops to all loop
orders for gauge theories in dimensions $3> d \geq 1$.

On the gravity side, we use the string duals for the sixteen
supercharge Yang-Mills theories proposed in \cite{Itzhaki:1998dd}.
These D$p$-brane geometries (where $d=p+1$) contain an $S^{8-p}$, the
$d$ boundary theory coordinates, and a $U$ direction, so that the
boundary is at $U=\infty$. We find the explicit fundamental string
solutions corresponding to circular Zarembo-type Wilson loops in these
backgrounds\footnote{String duals of generic Wilson loops have also
  been considered in the D$p$-brane geometries, see
  \cite{Sonnenschein:1999if}.}. They wrap part of an $S^2 \subset
S^{8-p}$ and extend in the $U$-direction from the boundary circle. We
find that these solutions have the expected zero regularized area.
This result is independent of the cut-off $U_{\text{max.}}$ where the
boundary theory is defined; this is the string-side manifestation of
the protection of these operators in the gauge theories, despite the
issues of running couplings and non-renormalizability. In appendix
\ref{sec:app} we analyze the supersymmetry respected by the solutions
and find that they are indeed $1/4$ BPS, as required. We also
generalize the framework of calibrated surfaces given in
\cite{Dymarsky:2006ve} to the D$p$-brane geometries, thereby proving
that the regularized action vanishes for any Zarembo-type Wilson loop
constructed in these theories, and as a check show that our circular
string solutions also satisfy the appropriate equations. Finally, in
section \ref{sec:zm} we discuss the potential string-side
manifestation of the gauge theory generalized Konishi anomaly for
$d=7$.

\section{Gauge theory results}
\label{sec:ss}

In this section we present the arguments in favor of the triviality of
the vacuum expectation values of supersymmetric Wilson loops in 16
supercharge super Yang-Mills theories, from the perspective of the
relevant gauge theories. To this end, we shall start with a
perturbative point of view, and subsequently correlate the
weak-coupling results with all-loop predictions based on superspace
techniques obtained in \cite{Guralnik:2003di}.

We start with a sixteen supercharge SYM action in $10>2\omega \geq 1$
dimensions given by \beq S = \frac{1}{g^2}\int d^{2\omega }x\, \Tr
\left( \frac{1}{2}F^2_{\mu \nu} + (D_\mu \Phi ^i)^2 -\frac{1}{2}[\Phi
  ^i,\Phi ^j]^2 + \bar{\Psi}\Gamma ^\mu D_\mu \Psi + i\bar{\Psi}\Gamma
  ^i[\Phi ^i,\Psi] \right). \eeq It is understood that the Lorentz
indices $\mu, \nu = 1,\ldots, 2\omega$ while the number of scalars $i,j
= 1,\ldots ,(10-2\omega )$.

As was shown by Zarembo in \cite{Zarembo:2002an}, the triviality of
the Wilson loop expectation value at the leading order in perturbation
theory is simply a consequence of the equality of the gluon and scalar
propagators in the Feynman gauge. Although the focus in
\cite{Zarembo:2002an} was on four dimensional gauge theory, this
leading order result readily generalizes to all the dimensional
reductions of the ten dimensional $\mathcal{N}=1$ gauge theory.

At the next-to-leading order, the diagrams that do not involve loop
corrections to propagators  cancel due to
the same reason as above.  In other words, the following cancelations
between Feynman diagrams occur for all dimensional reductions of
$\mathcal{N}=1$, $d=10$ SYM theories, due to the same arguments put
forward in the Feynman gauge for the four dimensional theory in
\cite{Zarembo:2002an}:
\begin{center}
\begin{tikzpicture}
\draw (0,0) circle (1cm);
\draw[snake=coil,segment length=4pt] (0,1) -- (0,-1);
\node at (1.5,0) [,] {$+$};
\draw (3,0) circle (1cm);
\draw (3,1) -- (3,-1);
\node at (4.5,0) [,] {$= 0,$};
\draw (0,-2.5) circle (1cm);
\draw[snake=coil,segment length=4pt] (-.5,-3.35) -- (-.5,-1.65);
\draw (.5,-3.35) -- (.5,-1.65);
\node at (1.5,-2.5) [,] {$+$};
\draw (3,-2.5) circle (1cm);
\draw (2.5,-3.35) -- (2.5,-1.65);
\draw (3.5,-3.35) -- (3.5,-1.65);
\node at (4.5,-2.5) [,] {$+$};
\draw (6,-2.5) circle (1cm);
\draw[snake=coil,segment length=4pt] (5.5,-3.35) -- (5.5,-1.65);
\draw[snake=coil,segment length=4pt] (6.5,-3.35) -- (6.5,-1.65);
\node at (7.5,-2.5) [,] {$=0,$};
\draw (0,-5) circle (1cm);
\draw[snake=coil,segment length=4pt] (0,-5) -- (0,-6);
\draw[snake=coil,segment length=4pt] (0,-5) -- (-.85,-4.5);
\draw[snake=coil,segment length=4pt] (0,-5) -- (.85,-4.5);
\node at (1.5,-5) [,] {$+$};
\draw (3,-5) circle (1cm);
\draw[snake=coil,segment length=4pt] (3,-5) -- (3,-6);
\draw (3,-5) -- (2.15,-4.5);
\draw (3,-5) -- (3.85,-4.5);
\node at (4.5,-5) [,] {$=0.$};
\end{tikzpicture}
\end{center}
For the triviality of the Wilson loop expectation value to hold at the
next-to-leading order, all that one needs to show is the equality
between the one loop corrected gluon and scalar propagators in the
Feynman gauge, such that the following cancelation takes place:
\begin{center}
\begin{tikzpicture}
\draw (0,0) circle (1cm);
\fill[black] (0,0) circle (.5cm);
\draw[snake=coil,segment length=4pt] (0,-1) -- (0,-.5);
\draw[snake=coil,segment length=4pt] (0,.5) -- (0,1);
\node at (1.5,0) [,] {$+$};
\draw (3,0) circle (1cm);
\fill[black] (3,0) circle (.5cm);
\draw (3,-1) -- (3,-.5);
\draw(3,.5) -- (3,1);
\node at (4.5,0) [,] {$=0.$};
\end{tikzpicture}
\end{center}
The one-loop gluon propagator in this gauge is given by \beq \Delta
^{ab}_{\mu\nu} = g^2\delta ^{ab}\frac{1}{p^2}\left(\delta _{\mu\nu} -
  g^2N\frac{\Gamma(2-\omega)\Gamma(\omega)\Gamma(\omega
    -1)}{(4\pi)^\omega\Gamma(2\omega)}f_g(\omega)\frac{\delta_{\mu\nu}-p_\mu
    p_\nu/p^2}{p^{4-2\omega}}\right),  \eeq where the function $f_g$ encodes
the contributions to the propagator from the various interaction
vertices  \beq f_g = 2(3\omega -1) - N_s -N_f(\omega -1). \eeq The
contribution of $2(3\omega -1)$ in $f_g$ is due to the combination of
the gluon-gluon and ghost-gluon scattering in $2\omega $
dimensions. The factor of $N_s$ arises from the $N_s$ real adjoint
scalars running in loops, while the factor of $N_f$; the number of
real fermionic degrees of freedom in the theory, is due to
gluon-fermion scattering.

Using the same notation, we may write the one loop corrected scalar
propagator as \beq \Delta ^{ab}_{mn} = g^2\delta
^{ab}\frac{1}{p^2}\left(\delta _{mn} -
  g^2N\frac{\Gamma(2-\omega)\Gamma(\omega)\Gamma(\omega
    -1)}{(4\pi)^\omega\Gamma(2\omega)}f_s(\omega)
\frac{\delta_{mn}}{p^{4-2\omega}}\right),
\eeq where \beq f_s(\omega ) = 4(2\omega -1) -
\frac{N_f}{2}(2\omega -1).  \eeq
The contribution of $4(2\omega -1)$ comes about due to the
scalar-vector intermediate state,
while the fermion loop contribution to the scalar propagator generates
a factor of $\frac{N_f}{2}(2\omega -1)$ with the opposite sign.\\

A necessary and sufficient condition for the supersymmetric Wilson
loops to have unit vacuum expectation value at the one and two loop
level is \beq f_g = f_s. \eeq It is easy to check that this is indeed
satisfied when the number of real scalars $N_s = 10-2\omega$ and
$N_f=16$.

We have thus established the triviality of the Wilson loop expectation
value at the next to leading order for {\it all} dimensional
reductions of the $\mathcal{N} =1$ ten dimensional SYM.

The one loop corrected gluon and scalar propagators, as they have been
expressed above, are also valid for the dimensional reduction of the
six and four dimensional $\mathcal{N} =1$ SYM theories as well. The
equality of the loop corrected propagators continues to hold if we use
either \beqs
N_f = 8, \hspace{.3cm} N_s = 6-2\omega \hspace{.3cm}\mbox{or}\\
N_f = 4, \hspace{.3cm} N_s = 4-2\omega.  \eeqs This fact proves the
triviality of Wilson loop expectation value for eight (four)
supercharge theories in dimensions less than or equal to five (three).

Thus,  the following table summarizes the balance between
the number of dimensions and the number of supersymmetries necessary
for Wilson loops to have trivial expectation values at the
next-to-leading order:
\begin{center}
\begin{tabular}{cc}
  \hline
  Number of Supercharges & Dimensions $\leq$ \\
  16 & 9 \\
  8 & 5 \\
  4 & 3 \\
  \hline
\end{tabular}
\end{center}
It is probably too optimistic to expect that all the gauge theories
listed above retain the triviality of the Zarembo loops to all orders
in perturbation theory. However, for the case of the sixteen
supercharge theories, lower dimensional superspace techniques were
successfully employed in \cite{Guralnik:2003di} to probe the all-loop
behavior of many of the gauge theories considered above. We shall
briefly review these techniques and compare the superspace results
with the perturbative computations reported above.

For the four dimensional theory, the starting point was a rewriting of
the action in a $\mathcal{N} =2$, $d=1$ superspace, coordinatized by
$t, \theta_\alpha, \bar{\theta}_\alpha$, where $\alpha = 1,2$ is an
$SU(2)$ index. The action for the four dimensional gauge theory was
shown to be \cite{Guralnik:2003di}\footnote{${\cal W}_\a = \bar D \bar
  D e^V D_\a e^{-V}$, see \cite{Guralnik:2003di} for further details.}
\bsp S = \frac{1}{g^2}&\int d^3x\,dt\,\Biggl[ \Tr
\left( \mathcal{W}_\alpha \mathcal{W}^\alpha \epsilon_{ijk}(\Phi
  _i\partial_j\Phi_k + \frac{2i}{3}\Phi_i \Phi_j\Phi_k)+
  cc\right)_{\theta \theta}\\& +
\Tr \left(\bar{\Omega}_ie^V\Omega_ie^{-V}\right)_{\theta \theta
  \bar{\theta}\bar{\theta}}\Biggr].
\end{split}
\ee In the quantum mechanical superspace, the three chiral superfields
$\Phi _i$ contain the spatial components of the gauge potential $A_i$
and three of the six real scalars\footnote{At the risk of abuse of
  notation, we denote both the chiral superspace fields as well as the
  real scalars by $\Phi$. We hope that the difference will be clear
  from the context.} $\Phi ^{i+3}$. The bottom component of the chiral
fields being given by $A_i + i\Phi ^{i+3}$. The temporal component
$A_0$ as well as $\Phi ^{7,8,9}$ are contained in the vector
superfield $V$. The superfields are also implicitly labeled by the
coordinates $x_i$, which are treated simply as auxiliary indices from
the quantum mechanical point of view. $\Omega $ is given by \beq
\Omega _i = \Phi _i + e^{-V}(i\partial_i - \bar{\Phi}_i)e^V.  \eeq One
of the main observations in the paper was that the Wilson loops of the
type considered in this paper could be thought of as elements of a
chiral ring from the lower dimensional superspace point of view. In
particular the equation of motion for these loops took on the form
\beq \Bigl\la \Tr\Bigl(
W(C,x)\,\epsilon_{ijk}\mathcal{F}_{jk}(x)\Bigr)\Bigr\ra_{\theta =
  \bar{\theta} = 0} = \mathcal{A}_i, \eeq where, \beq \mathcal{F}_{jk} =
\partial _j \Phi_k - \partial _k \Phi_j + i[\Phi_j, \Phi _k], \eeq and
where $W(C,x)$ is the untraced Wilson loop operator with a marked
point $x$ on the loop, and $\mathcal{A}_i$ is a possible anomaly term.
In the absence of the anomaly term, the loop equation implied shape
independence. In conjunction with the fact that the loop is an element
of the chiral ring, the shape independence yielded a trivial
expectation value of the loop. Note that this hinges upon the
three-dimensional epsilon symbol and for this reason is limited to
curves in $\bR^3$.

Similar arguments were also applied to sixteen supercharge Yang-Mills
theories in dimensions $3\leq d\leq 7$. The key to the generalization
was being able to write the action for the relevant gauge theories in
 a four supercharge $d-3$ dimensional superspace. It was further
shown that only in the case of the seven dimensional gauge theory does
one encounter a non-zero anomaly; this is the generalized Konishi anomaly.

Conjoining these superspace arguments with the evidence presented from
the weak coupling perturbation theory, we conclude that sixteen
supercharge SYM theories in dimensions $6 \geq d \geq 3$ possess
supersymmetric Wilson loops with trivial vacuum expectation values.

It is also worth noting that lower dimensional superspace methods were
also employed to analyze Wilson loops in SYM theories with 8
supercharges, and Wilson loops with trivial expectation values were
found in $4\geq d \geq 1$ dimensions in \cite{Guralnik:2004yc}. These
results are consistent with the perturbative results reported earlier
for the dimensional reductions of $\mathcal{N} =1$, $d=6$ SYM. The case
of the five dimensional Yang-Mills theory, suffers from a
non-vanishing anomaly, which was not seen in the perturbative
calculations we presented above.

In summary, the next to leading order perturbation theory and the
superspace arguments match up in the following cases:
\begin{center}
\begin{tabular}{cc}
  \hline
  Number of Supersymmetries & Dimensions \\
  16 & $3\leq d \leq 6$\\
  8 & $1 \leq d \leq 4$ \\
  \hline
\end{tabular}
\end{center}
In the case of sixteen supercharge theories, we also have a dual
gravity description available to us. In what follows, we reproduce and
generalize the results for this case using the dual gravity
picture. As the table above indicates, apart from the usefulness of
the gravity computation as non-trivial test of the gauge gravity
duality, can hope to shed some light on the non-perturbative behavior
of the Zarembo loops for the gauge theories for which the lower
dimensional superspace arguments do not exist, e.g. the case of $d=2$
SYM with sixteen supercharges.

\section{String duals and strong coupling results}

The string duals of the class of maximally supersymmetric Yang-Mills
theories were presented in \cite{Itzhaki:1998dd}. The holographic dual
of the $d=p+1$ dimensional gauge theory is given by the string
frame metric
\bsp\label{pbrane}
&ds^2 = \a' \left( \frac{U^{(7-p)/2}}{C_p} dx_\shortparallel^2
+ \frac{C_p}{U^{(7-p)/2}} \, dU^2 + C_p\, U^{(p-3)/2}\,d\O_{8-p}^2
 \right),\\
 &e^\phi = (2\pi)^{2-p} g^2
 \left(\frac{C_p^2}{U^{7-p}}\right)^{(3-p)/4},~
 C_p^2 = g^2 N \,2^{7-2p} \pi^{(9-3p)/2}\, \G\left(\frac{7-p}{2} \right),
\end{split}
\ee
where $g$ and $N$ are the bare coupling and the number of colours of
the dual Yang-Mills theory. There is also a $p$-form gauge potential
which depends only on the $U$ coordinate. These solutions are obtained
from the field theory limit of D$p$-brane solutions
\be
g^2 = (2\pi)^{p-2} \,g_s\, {\a'}^{(p-3)/2}=\text{fixed},\quad \a'\to 0,
\ee
where one can see that for $p>3$, the string coupling $g_s\to\infty$
which indicates a breakdown of the limit, in the sense that $\a'$
corrections are not suppressed and the decoupling of bulk modes is not
guaranteed. This is a reflection of the fact that the Yang-Mills
theories with $d=p+1>4$ are nonrenormalizable. As discussed in the
introduction, we are describing objects which are protected and
therefore we can trust our solutions in spite of this
breakdown. Indeed we will find that the regularized action of our
string solutions vanishes independently of the choice of cut-off
$U_{\text{max.}}$ - the coordinate dual to the boundary gauge theory
energy scale.

\subsection{Supersymmetric circular loops}

We present here string solutions corresponding to circular
supersymmetric Wilson loops in the background (\ref{pbrane}). We have
a natural lower bound of $p=1$, in order that the boundary has enough
dimensions to accommodate the circle, namely two, and a natural upper
bound of $p=7$, since, as we will see below, we will require at least
an $S^1$ to accommodate the coupling of the Wilson loop to the scalars
of the dual gauge theory. We have analyzed the supersymmetry of these
solutions in appendix \ref{sec:app}, where we show that they are 1/4
BPS.

We begin with the action of the fundamental string in Euclidean
conformal gauge, in the background (\ref{pbrane}). With the cases
$p<7$ in mind, we write
\bsp
dx_{\shortparallel}^2 = dr^2 + r^2 d\psi^2 + dx_{p-1}^2,\\
d\O_{8-p} = d\T^2 + \cos^2\T\,d\phi^2 + \sin^2\T \, d\O_{6-p}^2,
\end{split}
\ee
where $dx_{p-1}^2$ is a $p-1$ dimensional metric on $\bR^{1,p-2}$ or
$\bR^{p-1}$ (in the case $p=1$ we are forced to take the Euclidean
metric). Our solution ansatz is then
\bsp\label{ansatz}
\psi = \phi = \t, ~\t \in [0,2\pi],~~
r = r(U),~~ \T = \T(U),~~
dx_{p-1}^2 = d\O_{6-p}^2 = 0,
\end{split}
\ee
with which we can write the string action as
\be
S = \frac{C_p}{4\pi} \int_0^{2\pi} d\t \int d\s \,\Biggl[
\frac{U^{(7-p)/2}}{C_p^2} \left(r^2 + {r'}^2\right)
  +\frac{{U'}^2}{U^{(7-p)/2}}
+U^{(p-3)/2} \left( {\T'}^2 + \cos^2\T \right) \Biggr],
\ee
where prime denotes differentiation w.r.t. $\s$. We must also satisfy
the Virasoro constraint
\be\label{virr}
\frac{U^{(7-p)/2}}{C_p^2}\, {r'}^2
  +\frac{{U'}^2}{U^{(7-p)/2}} +U^{(p-3)/2} {\T'}^2=
U^{(p-3)/2} \cos^2\T +\frac{U^{(7-p)/2}}{C_p^2}\, {r}^2.
\ee

The solution we find is
\bsp\label{sol}
&R^2-r(U)^2 =
\begin{cases}
\frac{2\,C_p^2}{5-p} \,U^{p-5},~ &p \neq 5\\
-2 \,C_5^2 \,\log U,~&p=5
\end{cases},\\
&\sin\T = \frac{U_{\text{min.}}}{U}, \qquad r(U_{\text{min.}})=0,
\end{split}
\ee
where $R$ is the asymptotic radius of the circle at $U=\infty$. Note
that for $p>4$, $r(\infty) = \infty$, $R$ becomes imaginary, and so
the solution doesn't satisfy the usual boundary condition. We will cut
the geometry off at $U_{\text{max.}}$ however, and so define the
radius of the circle in the boundary theory as $r(U_{\text{max.}})$.

The case $p=7$ is special, as $\theta=\theta'=0$ for this case; the
string wraps only the $S^1$ defined by the angle $\phi$. In order to
simplify the presentation we give the explicit solution in terms of
$\sigma$ here, and then continue with the cases $1 < p < 7$ below. The
solution for $p=7$ is
\be
U = U_{\text{min.}} \cosh \s, \quad r = C_7 \,U_{\text{min.}} \sinh \s,
\quad \s \in [0,\s_{\text{max.}}],
\ee
and leads to the action $S = C_7\, U_{\text{min.}}^2
\cosh\s_{\text{max.}} \sinh \s_{\text{max.}}$, which upon
regularization as in (\ref{legendre}) below, yields zero. 

Continuing with cases $1 < p < 7$, the solution (\ref{sol}) wraps one
half of an $S^2 \subset S^{8-p}$; the string worldsheet's boundary
lies along the equator\footnote{More precisely, when $U_{\text{max.}}$
  is not strictly $\infty$, the boundary is shifted down towards the
  pole.}.  In order to check that (\ref{sol}) is in fact a solution to
the equations of motion we express $r'$ and $\T'$ in terms of $U'$ and
plug them into the Virasoro constraint (\ref{virr}) and solve for $U'$
in terms of $U$.  The result of this operation is
\be\label{Upr}
U' = \begin{cases} \sqrt{\frac{2}{5-p}
   \left( U^{5-p}/U_{\text{min.}}^{5-p} -1\right)
    \left(U^2 - U_{\text{min.}}^2 \right) },~~&p\neq 5\\
  \sqrt{2  \log (U /
      U_{\text{min.}})\, \left(U^2 - U_{\text{min.}}^2
    \right) },~~&p=5.
\end{cases}
\ee
With this expression we can also express $U''$ in terms of $U$,
and through (\ref{sol}), we can therefore also express $r''$ and
$\T''$ in terms of $U$. The expression for $U''$ is
\bsp
&U'' = \frac{U_{\text{min.}}^{p-5}}{5-p} \Bigl(
(5-p)\,U^{4-p}\,(U^2-U_{\text{min.}}^2) + 2 U (U^{5-p}-U_{\text{min.}}^{5-p})
\Bigr),\quad p\neq 5,\\
&U'' =
U^{-1}\,(U^2-U_{\text{min.}}^2) + 2 U \log (U/ U_{\text{min.}})
,\quad p = 5.
\end{split}
\ee
It is then a straightforward, if somewhat tedious exercise to verify
that the equations of motion for $U$, $r$, and $\T$ are satisfied
through the chain of substitutions.

\begin{figure}\label{fig:sols}
\includegraphics*[bb= 0 0 685 430,width=6in]{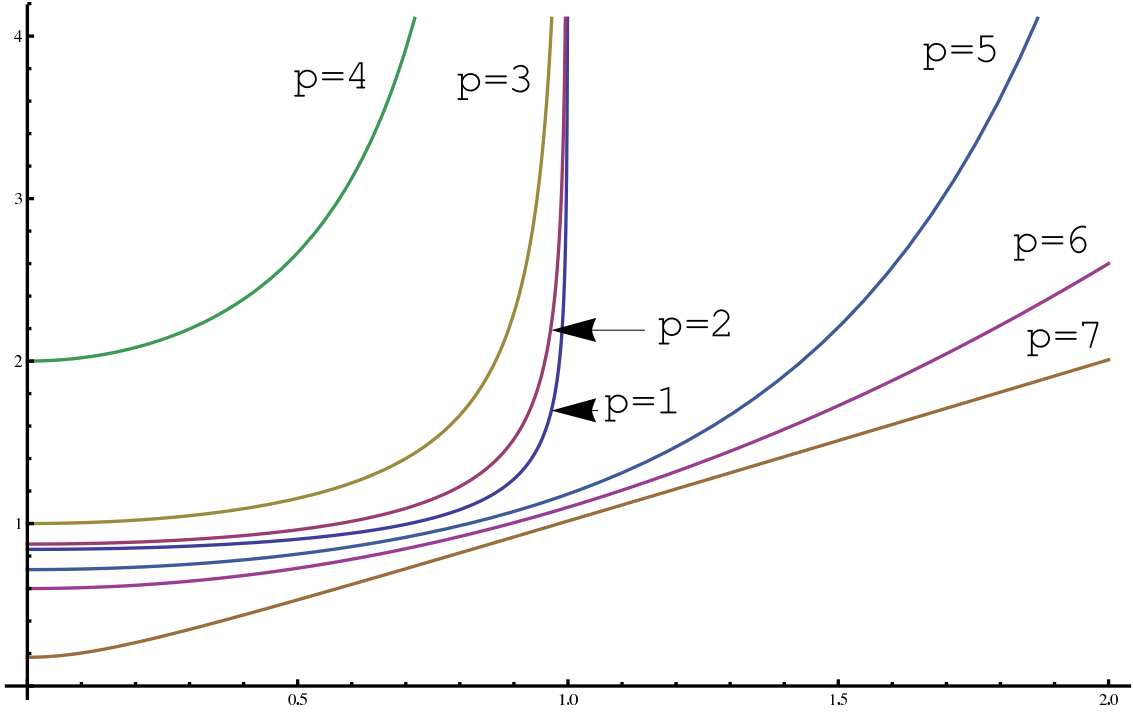}
\caption{A plot of $U$ vs. $r$ for the solutions (\ref{sol}). We have
  set $C_p=1$ and chosen $R$ differently in each case to accommodate
  the plots.}
\end{figure}

We have plotted $U(r)$ in figure \ref{fig:sols}.

It remains to compute the action of the solutions. Using (\ref{virr})
we can express the action as twice the ``prime'' terms, i.e. those
involving derivatives by $\s$. We express everything in terms of $U$
and $U'$, the latter we use to reexpress the integration over $\s$ by
integration over $U$.
\bsp
&S = C_p \int_{U_{\text{min.}}}^{U_{\text{max.}}}\frac{ dU}{
U^{(7-p)/2}} \, \frac{  \frac{2}{5-p}\,U^2
   \left( U^{5-p}/U_{\text{min.}}^{5-p} -1\right)
    + U^2 - U_{\text{min.}}^2}{\sqrt{\frac{2}{5-p}
   \left( U^{5-p}/U_{\text{min.}}^{5-p} -1\right)
    \left(U^2 - U_{\text{min.}}^2 \right)}},~~p\neq 5, \\
&S = C_5 \int_{U_{\text{min.}}}^{U_{\text{max.}}}\frac{ dU}{
U} \, \frac{  2 U^2 \log(U/U_{\text{min.}})
    + U^2 - U_{\text{min.}}^2}{\sqrt{  2\log (U /
      U_{\text{min.}})\, \left(U^2 - U_{\text{min.}}^2
    \right) }},~~p=5.
\end{split}
\ee
The integral is simple to evaluate. The result is
\bsp\label{Sres}
S = U_{\text{max.}}  \sqrt{1 - U_{\text{min.}}^2/U_{\text{max.}}^2}
\,r(U_{\text{max.}})  .
\end{split}
\ee
The prescription for removing the divergence from the action is to
perform a Legendre transformation \cite{Drukker:1999zq}, as follows
\bsp\label{legendre}
S_{\text{reg.}} &= S - \int d\t\, d\s \,
\p_\s \left({\cal Y}^I \frac{\d S}{\d \p_\s {\cal Y}^I} \right)\\
& = S - \int d\t\, {\cal Y}^I \frac{\d S}{\d \p_\s {\cal Y}^I} \Biggr|_{\p\S},
\end{split}
\ee
where we are using the coordinates defined in (\ref{Ycoords}). We then
find, using (\ref{Yembed}) and (\ref{Upr}),
\be
\frac{\d S}{\d \p_\s {\cal Y}^I}  = \frac{C_p}{2\pi} \,U^{(p-7)/2}\,\left(U \hat\T^I
\right)' = \frac{1}{2\pi} \,r(U)\, \Bigl(-\sin\t,\cos\t,0,\ldots,0\Bigr),
\ee
and so
\be
\int d\t\, {\cal Y}^I \frac{\d S}{\d \p_\s {\cal Y}^I} \Biggr|_{\p\S}
= U_{\text{max.}}  \sqrt{1 - U_{\text{min.}}^2/U_{\text{max.}}^2}
\,r(U_{\text{max.}}) .
\ee
We therefore have that
\be
S_{\text{reg.}} = 0 ,
\ee
independent of $U_{\text{max.}}$ and consistent with our expectations.

\subsection{Calibrated surfaces}
\label{sec:cal}

In the paper \cite{Dymarsky:2006ve} a method of calibrated surfaces
was employed to prove that the string duals of the supersymmetric
Wilson loops of general shape in the $p=3$ case had the expected
regularized action, namely zero. We now show that this machinery
applies equally well to the case of general $p$. As a check on our
work, we also show that it applies to the solutions (\ref{sol}).

In order to apply the technique we express the metric of $S^{8-p}$
together with the $dU^2$ term from (\ref{pbrane}) as follows
\be \label{Ycoords}
\frac{dU^2}{U^2}+d\O_{8-p}^2 = \frac{d{\cal Y}^I d{\cal
    Y}^I}{{\cal Y}^2},\quad {\cal Y}^I = U \hat \T^I, \quad \hat \T^I
\hat \T^I=1, \quad I=1,\ldots,9-p.  \ee
For convenience we will rescale the $x_{\shortparallel}^\m = C_p
X^\m$. We then have
\be\label{calmet}
ds^2 = \a' \,C_p \, \Bigl(
{\cal Y}^{(7-p)/2} dX^\m dX^\m + {\cal Y}^{(p-7)/2} d{\cal Y}^I d{\cal Y}^I
\Bigr).
\ee
Now we make a split in the ${\cal Y}^I$ coordinates
\be
{\cal Y}^I = (Y^m, V^i) ,\quad m=1,\ldots,p+1,\quad i = 1,\ldots,8-2p,
\ee
so that $Y^m$ and $X^\m$ have the same number of components. It is
clear that this can only be done for $p\leq 4$. For $p>4$ we can
choose instead to split the $X^\m = (X^{\bar I},V^i)$, so that
$X^{\bar I}$ has the same number of components as ${\cal Y}^I$, and what
follows is equally true (with the appropriate relabelling of indices).
We use precisely the same definition for an almost complex structure
proposed in \cite{Dymarsky:2006ve}
\be
J = J_{AB} \, d\bX^A \w d\bX^B = \d_{\m m} \, dX^\m \w dY^m,
\ee
where $\bX^A = (X^\m,Y^m,V^i)$. We find that the following key
relations used in \cite{Dymarsky:2006ve} are equally true for the
metric (\ref{calmet}), namely
\bsp\label{calrel}
&J_A^B\,J_B^C = -\d_A^\m \,\d_\m^C - \d_A^m \d_m^C,\\
&G_{MN} J^M_\m J^N_\n = G_{\m\n},\quad
G_{MN} J^M_m J^N_n = G_{mn},\quad
G_{MN} J^M_i J^N_j = 0.
\end{split}
\ee
That being the case everything follows as in
\cite{Dymarsky:2006ve}. We continue by reiterating the results of
\cite{Dymarsky:2006ve} in the interest of readability. One defines
\bsp
&{\cal P} \equiv \frac{1}{4} \int d^2\s \sqrt{h}\, h^{ab} \,
G_{MN} \, v^M_a v^N_b,\\
&v^M_a \equiv \p_a \bX^M - J^M_N \, \frac{h_{ac} \e^{cb}}{\sqrt{h}} \,\p_b \bX^N,
\end{split}
\ee
where $h_{ab}$ is a positive definite metric on the worldsheet. Using
(\ref{calrel}) one can then show that
\be\label{prel}
{\cal P} = \frac{1}{2} \int d^2\s \sqrt{h}\, h^{ab} \,
G_{MN} \, \p_a \bX^M \p_b \bX^N
- \int_{\S} J - \frac{1}{4} \int d^2\s \sqrt{h}\,h^{ab}
G_{ij}\,
\p_a V^i \p_b V^j,
\ee
where $\S$ is the string worldsheet. Now suppose that $v_a^M = 0$. As
can be easily checked in conformal gauge, this condition automatically
implies that the string equations of motion and Virasoro constraints
are satisfied. Further, this implies that ${\cal P} = 0$ and that
$\p_a V^i=0$. We then have that
\be
S = \frac{C_p}{2\pi} \int_\S J,
\ee
that is, we have that the action of the string worldsheet is
expressible as an integral of the closed 2-form $J$ over the string
worldsheet. This will integrate to a surface term
\be\label{Jchain}
\int_{\S} J = \int_{\S} \d_{\m m}\, d ( Y^m dX^\m ) =
U_{\text{max.}} \int_{\p\S} \d_{\m m}\,\hat \T^m \,dX^\m.
\ee
Now examining the equation $v_a^M =0$ in conformal gauge one finds
\be
\dot X^\m = U^{(p-7)/2}\, {Y'}^m
=U^{(p-7)/2} \Bigl( U \,({\hat \T}^m)' + \hat \T^m \, U'\Bigr).
\ee
If it is true that $U^{(p-5)/2}\,({\hat \T}^m)'\to 0$ as the boundary is
approached, one then has that
\be\label{Tbc}
\dot X^\m(U_{\text{max.}}) \simeq  U_{\text{max.}}^{(p-7)/2}
(\hat \T^m \, U')|_{U_{\text{max.}}},
\ee
and so
\be
\hat \T^m|_{\p \S} = \frac{\dot x^\m}{|\dot x|},
\ee
where $x^\m = X^\m(U_{\text{max.}})$ is the Wilson loop contour. This
is precisely the contour of the supersymmetric Wilson loop in the
gauge theory (\ref{pc}), (\ref{gbc}). Thus we have found a solution to
the string equations of motion which also satisfies the necessary
boundary conditions. Furthermore, in conformal gauge we have that
\be
U^{(p-7)/2}\, {Y'}^m = \frac{2\pi}{C_p}\frac{\d S}{\d \p_\s Y^m} ,
\ee
and therefore (\ref{Jchain}) also gives
\be
S = \oint d\t\, Y^m\,\frac{\d S}{\d \p_\s Y^m}\Biggr|_{\p\S} ,
\ee
which is nothing but the divergence removed from the action by the
Legendre transformation \cite{Drukker:1999zq} to give
$S_{\text{reg.}}$; thus we see that the regularized action vanishes
for these solutions. Again this result is independent of the choice of
cut-off in the coordinate $U$.

\subsubsection{Checking the circular supersymmetric solutions}

We can now verify that our solution (\ref{sol}) obeys the equations
$v^M_a = 0$ and (\ref{Tbc}), thereby confirming our result
(\ref{Sres}). We begin by writing our solution (\ref{sol}) in the
coordinates (\ref{calmet}). We find (for example, for $p\neq
5,\,7$)\footnote{The $p=5,\,7$ cases follow similarly.}
\bsp\label{Yembed}
&X^1 = r\cos\psi = \sqrt{\frac{2}{5-p} \left(U_{\text{min.}}^{p-5}
- U^{p-5}\right)} \cos \t,\\
&X^2 = r\sin\psi = \sqrt{\frac{2}{5-p} \left(U_{\text{min.}}^{p-5}
- U^{p-5}\right)} \sin \t,\\
&Y^1 = -U \cos\T \sin\phi = - \sqrt{U^2-U_{\text{min.}}^2} \sin\t,\\
&Y^2 = U \cos\T \cos\phi =  \sqrt{U^2-U_{\text{min.}}^2} \cos\t,\\
&Y^3 = U\sin\T = U_{\text{min.}}.
\end{split}
\ee
The equations $v^M_a = 0$ in conformal gauge then reduce to
\bsp
{X'}^\m + \left(Y^2 + V^2\right)^{(p-7)/4} \dot Y^{m=\m} = 0,\\
{Y'}^m - \left(Y^2 + V^2\right)^{(7-p)/4} \dot X^{\m=m} = 0,\\
\dot X^\m -  \left(Y^2 + V^2\right)^{(p-7)/4} {Y'}^{m=\m}=  0,\\
\dot Y^m + \left(Y^2 + V^2\right)^{(7-p)/4} {X'}^{\m=m}=  0,
\end{split}
\ee
which, through use of (\ref{Upr}) may be shown to be
satisfied. Finally we note that
\bsp
&U^{(p-5)/2}\,({\hat \T}^m)' = \frac{U^2_{\text{min.}}}{U^2} \, r(U)
\,\Bigl(-\sin\t,\cos\t,0,\ldots,0\Bigr),\\
&U^{(p-7)/2} \hat \T^m \, U' =
\left(1-\frac{U_{\text{min.}}^2}{U^2}\right)r(U)
\,\Bigl(-\sin\t,\cos\t,0,\ldots,0\Bigr),
\end{split}
\ee
and so (\ref{Tbc}) is also satisfied.

\subsection{Zero modes and the generalized Konishi anomaly in $d=7$}
\label{sec:zm}

It seems a contradiction that in the gauge theory analysis discussed
in section \ref{sec:ss} there is an anomaly in the case $d=7$
precluding $\la W\ra =1$ for this theory, whereas the string solution
seems to suffer no such issue. In fact, as discussed in the
introduction, there is more to the vacuum expectation value of the
Wilson loop than the regularized action; the prefactor ${\cal V}$
stemming from integration over zero modes in the partition function
also plays a role. The issue, as regards supersymmetric Wilson loops,
was first discussed in \cite{Zarembo:2002an}, for the case
$p=3$. Although there appears to be no parametric freedom for the
minimal area embedding of a string in an $AdS$ space with given
boundary conditions, Zarembo argued that the supersymmetric circle may
be embedded into the $S^5$ with a freedom given by a vector ${\bf n} \in
S^3$ which chooses which $S^2$ the worldsheet occupies. This gives a
natural reason for the cancellation of the prefactor in $\la W \ra$,
as these three zero modes could cancel the effect of the basic three
coming from the $AdS$ embedding. This reasoning is limited to the case
of planar curves, and \cite{Dymarsky:2006ve} noted the lack of
resolution of this problem for general curves. Specifically, in order
for the $R$-symmetry of the Wilson loop defined in the gauge theory to
match the string solution, these zero modes must be integrated over.

In our case we note the fact that uniquely in the case of $d=7$ (i.e.
$p=6$) do we have that the spherical product space $S^{8-p}$ is an
$S^2$. In this case we are restricted to curves in $\bR^3$, and we
will concentrate on our explicit solution for the supersymmetric
circle, although we expect the following comments to be true for
general closed curves in $\bR^3$. The fact that the spherical product
space is an $S^2$ precludes the existence of zero modes on the
spherical side of the geometry, and thus, assuming the absence of any
parametric freedom in the embedding on the analogue of the $AdS$ side
of the geometry, precludes the possible cancellation of the three
basic zero modes of the string worldsheet. We would thus expect a
non-zero prefactor ${\cal V} \sim \lambda^{-3/4}$ and therefore our
prediction for the vacuum expectation value of the Wilson loop at
strong coupling is
\be\label{d7koni}
\la W \ra_{d=7} \sim \left(\frac{\lambda}{R^3}\right)^{-3/4},
\ee
where $R$ is a scale setting the size of the Wilson loop. This seems
to be the string-side manifestation of the generalized Konishi anomaly
in $d=7$ discussed in section \ref{sec:ss}. It would be very
interesting to try to recover this result from gauge theory.

We also note that for $d=8$, we have Wilson loops in $\bR^2$ described
by a string worldsheet wrapping a spherical product space which is
simply an $S^1$. Here too we do not expect parametric freedom in the
string embedding, and so again would expect the basic three zero modes
to give $\la W\ra \neq 1$, as in (\ref{d7koni}) above. It would, of course, be of great
interest to uncover a gauge theoretic mechanism for this potential anomaly in $d=8$.

The situation is extremely reminiscent of the circular Wilson loop for
$\mathcal{N}=d=4$ SYM obtained by a ``large'' conformal transformation
of the straight line. In that case, the Wilson loop expectation value
is a non-trivial function of the 't Hooft coupling. However, the
vacuum expectation value for the loop is entirely determined by an
anomaly; namely, the conformal anomaly \cite{Drukker:2000rr}. For the
seven dimensional gauge theory, the generalized Konishi anomaly seems
to play a similar role. It is tempting to speculate that it might
similarly be possible to recover the strong coupling result mentioned
above from the gauge theory end, by reducing the problem to a matrix
model computation\footnote{To this end, it might be interesting to
  investigate if the methods of \cite{Bonelli:2008rv} can be adapted to
  the analysis of the gauge theory in $d=7$.}.

\section{Summary and outlook}

In this paper we have generalized the construction of supersymmetric
Wilson loops in ${\cal N}=4$, $d=4$ SYM at weak and strong coupling to
the general case of SYM theories with 16 supercharges in $d$
dimensions (and in the case of $d\leq 4$ ($d\leq 3$) at weak coupling,
with 8 (4) supercharges). We have given two-loop perturbative evidence
and reviewed the applicability of evidence from superspace techniques,
that these loops have trivial vacuum expectation values. Using the
gauge/strings duality we have also described the 16 supercharge theory
supersymmetric Wilson loops at strong coupling and also found strong
evidence of trivial expectation values; the dual string solutions have
zero regularized action. We have found the explicit fundamental string
solutions for the case of circular supersymmetric loops in general
$d$. In the case of $d=7$ where superspace techniques indicate a
non-zero expectation value on the gauge theory side, we have found a
strong candidate dual manifestation of this phenomena at strong
coupling, namely the disappearance of string worldsheet zero modes.
Based on this we have given a prediction for the strong coupling
behavior of the vacuum expectation value of supersymmetric Wilson
loops in the $d=7$ theory.

Looking beyond the issues addressed in this paper, it would doubtless
be interesting to try and extend the present results to more general
Wilson loops and to other instances of gauge/gravity dualities. For
example, for the theories we considered with $d>4$, the various UV
completions were discussed in \cite{Itzhaki:1998dd} (see also
\cite{Sonnenschein:1999if}) and involve lifting to M-theory (in the
case of odd $d$), or the application of S-duality in the IIB case. We
expect these theories to retain the trivial Wilson loop operators we
have constructed here as a natural consequence of coupling
independence. On a different note, it was pointed out earlier in the
paper, using both perturbative as well as superspace methods, that
eight supercharge SYM theories in $1\leq d\leq 4$ admit Zarembo
loops. Clearly, this fact can be used to carry out non-trivial tests
for any candidate gravity dual for these theories.

In the special case of three spacetime dimensions, the recent
developments due to Bagger, Lambert and Gustavsson (BLG)
\cite{Bagger:2006sk,Bagger:2007jr,Bagger:2007vi,Gustavsson:2007vu} and
Aharony, Bergman, Jafferis and Maldacena (ABJM) \cite{Aharony:2008ug}
relate the sixteen supercharge SYM theory to superconformal
Chern-Simons (SCS) theories. The $\mathcal{N} = 8$ SCS theory proposed
by BLG, which can also be recovered as a special case of the ABJM
model, is believed to be related to the IR limit of the sixteen
supercharge SYM theory. It is interesting to note that Wilson loops
that preserve global supersymmetries have also been constructed for
the ABJM model in \cite{Berenstein:2008dc,Drukker:2008zx, Chen:2008bp,
  Rey:2008bh}. In the present paper, we have shown that Zarembo loops
exist in the SYM theory both at weak and at strong coupling. It thus
seems plausible that one can uncover a precise relationship between
the Zarembo loops of the SYM theory and the corresponding operators in
the BLG model. Perhaps, the formal relationship between the
Lagrangians of the two theories, elucidated in \cite{Mukhi:2008ux},
can prove to be fruitful to uncover this aspect of the M2/D2 duality.

On a related note, it would be extremely interesting to explore
connections between Wilson loops and scattering amplitudes. The
relations between these two classes of gauge theory observables, first
studied in the context of $\mathcal{N}=4 $ SYM in $d=4$
\cite{Alday:2007hr,
  Drummond:2007aua,Brandhuber:2007yx,Alday:2007mf,Alday:2007he} can
potentially exist for three dimensional Yang-Mills theories as
well. The matrix structure of all $2\leftrightarrow 2$ scattering
amplitudes for ${\cal N}\geq 4$ SCS theories was recently explored in
\cite{Agarwal:2008pu}. This study includes the BLG model, which is
expected to be the strong coupling dual of the sixteen supercharge SYM
theory. A further study of Wilson loops in the three dimensional gauge
theory is obviously needed to fill the missing connections between
Wilson loops and scattering amplitudes both in the SYM theory as well
as its dual strong-coupling description as a SCS theory.

\section*{Acknowledgements}

D.Y. acknowledges the support of the Natural Sciences and Engineering
Research Council of Canada (NSERC) in the form of a Postdoctoral
Fellowship, and also support from the Volkswagen Foundation. A.A.
wishes to thank Tristan McLoughlin for interesting discussions about
this work.


\appendix

\section{Supersymmetry of string solutions}
\label{sec:app}

The supersymmetry analysis of the case $p=3$ is given in
\cite{Drukker:2006ga}, and is rather special due to the constancy of
the dilaton. Therefore here we will present the analysis for the cases
$p\neq 3$.

The Killing spinor equations for the geometries (\ref{pbrane}) are
obtained by demanding that the variation of the dilatino $\l$ and
gravitino $\psi_M$ vanish on the supergravity solution. We use the
``democratic formalism''\footnote{In the democratic formalism the
  number of Ramond-Ramond potentials $C_{(n)}$ is doubled so that
  $n=0,2,\ldots,10$ for IIB and $n=1,3,\ldots,9$ for IIA. The extra
  potentials, in the absence of fermionic and NS-NS fields, are simply
  given by the action of Hodge duality upon the field strengths.}
developed in \cite{Bergshoeff:2001pv} (see for example
\cite{Koerber:2007hd}, appendix B therein),
\bsp\label{gravdil}
\d \psi_M = D_M \,\e + \frac{e^\phi}{16}\, \wt F_{(p+2)} \,\g_M
\,{\cal P}_{p}\, \e = 0,\\
\d \l = \wt \p \phi \, \e + \frac{e^\phi}{8}\, (-1)^p\,\wt
F_{(p+2)}\,{\cal P}_{p}\, \e = 0,
\end{split}
\ee
where we have used the fact that the $p$-brane solutions have only a
dilaton $\phi$ and a $(p+2)$-form field strength $F_{(p+2)}$ turned on.
The Killing spinor is denoted by $\e$ while $\g_M$ are the real 10-d
curved space gamma matrices in Lorentzian mostly positive signature.
The covariant derivative $D_M = \p_M + \frac{1}{4} \o_M^{ab} \G_{ab}$
where $\G_a$ denote tangent space gamma matrices. The constant
matrices ${\cal P}_{p}$ are given in \cite{Koerber:2007hd} but won't
concern us here. Finally we have adopted the notation
\be
\wt F_{(p+2)} \equiv F_{M_1 \ldots M_{p+2}} \g^{M_1 \ldots
  M_{p+2}},\quad
\wt \p \phi \equiv \g^M \p_M \phi.
\ee
By acting with $\g_M$ from the left on the second equation in
(\ref{gravdil}) we may eliminate the field strength term in the first
equation and obtain
\be\label{kseq}
D_M \,\e -\frac{1}{2}\frac{s_M}{(3-p)}\, \wt \p \phi\, \g_M \, \e = 0,
\quad s_M = \begin{cases}
 1 ~~\text{if}~~M=0,\ldots,p\\
-1 ~~\text{otherwise}
\end{cases},
\ee
where we have used the fact that on the solution (\ref{pbrane}) the
dilaton $\phi$ depends only on the coordinate $U$, while $F_{(p+2)} =
F_{0\ldots p\, U}$ where $0,\ldots,p$ denote the $p+1$ coordinates
$x_\shortparallel$.

For convenience we scale the $C_p$ and $\a'$ dependence out of the
metric, which is equivalent to replacing $\a',C_p\to 1$. We also
specialize to those coordinates relevant to the string solution
(\ref{ansatz}).  We then employ the following basis of one-forms
\bsp
&e^{\bar U} = U^{(p-7)/4} \, dU,\quad e^{\bar r} = U^{(7-p)/4} \, dr,
\quad e^{\bar\psi} = r U^{(7-p)/4} \, d\psi,\\
&\qquad\qquad e^{\bar\T} = U^{(p-3)/4} \, d\T,\quad
e^{\bar\phi} = U^{(p-3)/4} \cos\T \, d\phi,
\end{split}
\ee
using which the relevant components of the spin-connection are
\bsp
&\o^{\bar U \bar r}_r = \frac{p-7}{4} U^{(5-p)/2},\quad
\o^{\bar U \bar \psi}_\psi = \frac{p-7}{4}\, r\,  U^{(5-p)/2},\quad
\o^{\bar r \bar \psi}_\psi = -1,\\
&\qquad \o^{\bar U \bar \T}_\T = \frac{3-p}{4},\quad
\o^{\bar U \bar \phi}_\phi = \frac{3-p}{4} \cos \T,\quad
\o^{\bar \T \bar \phi}_{\phi} = \sin \T.
\end{split}
\ee
The Killing spinor equations (\ref{kseq}) are then given by
\bsp
&\p_U\,\e + \frac{p-7}{8U} \, \e = 0,\\
&\p_r \, \e = 0,\\
&\p_\psi\,\e - \frac{1}{2} \G_{\bar r \bar \psi} \, \e = 0,\\
&\p_\T \, \e - \frac{1}{2} \G_{\bar U \bar \T} \, \e = 0,\\
&\p_\phi \, \e + \frac{1}{2} \sin\T\, \G_{\bar \T \bar \phi} -
\frac{1}{2} \cos \T \,\G_{\bar U \bar \T} \, \e = 0,
\end{split}
\ee
and solved by
\be
 \e = U^{(7-p)/8} \,
e^{\frac{\T}{2} \G_{\bar U \bar \T}}\,
e^{\frac{\psi}{2} \G_{\bar r \bar \psi}}\,
e^{-\frac{\phi}{2} \G_{\bar\phi \bar U}} \, \e_0.
\ee

The supersymmetry projector for the fundamental string is given by
\be
\p_\t \bX^M \p_\s \bX^N \, \g_{MN} \,\e = \sqrt{-\det \p_a \bX^M \p_b
  \bX^N G_{MN} } \,{\cal P}\, \e  = {\cal L}{\cal P}\, \e\,,
\ee
where
\be
{\cal P} = \begin{cases}
\G_{0}\ldots \G_{9},\quad &\text{IIA, i.e. $p$ even}\\
K I,\quad &\text{IIB, i.e. $p$ odd}
\end{cases},
\ee
where $KI=-IK$, $K$ acts by complex conjugation upon spinors while $I$
acts as $-i$, see
\cite{Skenderis:2002vf,Cederwall:1996ri,Cederwall:1996pv}. On our
solution (\ref{sol}) we find
\bsp
\p_\t \bX^M \p_\s \bX^N \, \g_{MN} =
U' \, r \, \G_{\bar\psi\bar U} + U'\,U^{(p-5)/2}\, \cos \T\, \G_{\bar
  \phi \bar U}\\
+ r' \, r \, U^{(7-p)/2} \, \G_{\bar\psi\bar r}
+ r' \, U \, \cos\T \,\G_{\bar \phi \bar r}\\
+ \T' \, U \, r \,\G_{\bar \psi \bar \T} + \T' U^{(p-3)/2} \cos \T
\,\G_{\bar \phi \bar \T}.
\end{split}
\ee
The Killing spinor also simplifies to
\be
\e = U^{(7-p)/8} \,
e^{\frac{\T}{2} \G_{\bar U \bar \T}}\,
e^{\frac{\t}{2} \left(\G_{\bar r \bar \psi}
- \G_{\bar\phi \bar U}\right)}
\, \e_0.
\ee
In order to find solutions to the projector equation, we find that we
must remove the $\t$ dependence from the Killing spinor by requiring
\be\label{bps1}
\G_{\bar r \bar \psi} \, \e_0 =
\G_{\bar \phi \bar U} \, \e_0.
\ee
The projector equation is then
\bsp
e^{-\frac{\T}{2} \G_{\bar U \bar \T}}\,
\Biggl[U' \, r \, \G_{\bar\psi\bar U} + U'\,U^{(p-5)/2}\, \cos \T\, \G_{\bar
  \phi \bar U}
+ r' \, r \, U^{(7-p)/2} \, \G_{\bar\psi\bar r}
+ r' \, U \, \cos\T \,\G_{\bar \phi \bar r}\\
+ \T' \, U \, r \,\G_{\bar \psi \bar \T} + \T' U^{(p-3)/2} \cos \T
\,\G_{\bar \phi \bar \T}\Biggr]\,
e^{\frac{\T}{2} \G_{\bar U \bar \T}} \, \e_0 = {\cal L} {\cal P} \e_0.
\end{split}
\ee
Expanding out the LHS of this expression and using (\ref{bps1})
one finds
\bsp
&-\Bigl( \sin\T \,r\,U' + \cos\T \, \T'\,r\,U \Bigr)\,\G_{\bar\T \bar
  \psi}\,\e_0
-\Bigl( \sin\T \cos\T\, U' \, U^{(p-5)/2} + \T' \, U^{(p-3)/2} \,\cos^2\T
\Bigr)\,\G_{\bar\T \bar \phi}\,\e_0\\
&+\Bigl( U'\, U^{(p-5)/2} \, \cos^2\T - r' \, r \, U^{(7-p)/2}
- \T' \, \sin\T\,\cos\T\,U^{(p-3)/2} \Bigr) \, \G_{\bar\phi \bar U}\,\e_0\\
&+\Bigl(U'\,r\,\cos\T + r'\,U\,\cos\T - \T' \, r\, U\,\sin\T \Bigr)
\,\G_{\bar\psi \bar U}\,\e_0.
\end{split}
\ee
One then finds that the first three bracketed expressions are zero on
the solution (\ref{sol})\footnote{The analysis is also valid for
  $p=7$, where we note that $\theta = \theta' = 0$.}, while the last
bracketed expression is equal to $\sqrt{\det \p_a \bX^M \p_b \bX^N
  G_{MN} }$, which by the Virasoro constraint (\ref{virr}) is the
square-root of a perfect square. In addition to (\ref{bps1}) we
therefore also have
\be\label{bps2}
\G_{\bar \psi \bar U} \, \e_0 = i {\cal P} \,\e_0.
\ee
The two conditions (\ref{bps1}) and (\ref{bps2}) each reduce the
supersymmetry by half, thus the solutions respect a quarter of the
original 16 supersymmetries, i.e. they are 1/4 BPS.

\bibliography{final}
\end{document}